\documentclass[11pt,aps,preprint]{revtex4}
\usepackage[active]{srcltx}
\usepackage[utf8]{inputenc}
\usepackage{latexsym}
\usepackage{amsmath}
\usepackage{graphicx}
\def\der{{\buildrel\,\leftrightarrow\over\partial}\!\!}

\begin{document}

\title{On Ward Identities in Lifshitz-like Field Theories}

\author{Pedro R. S. Gomes}
\email{pedrorsg@fma.if.usp.br}
\affiliation{Instituto de F\'\i sica, Universidade de S\~ao Paulo\\
Caixa Postal 66318, 05315-970, S\~ao Paulo, SP, Brazil}%
\author{M. Gomes}
\email{mgomes@fma.if.usp.br}
\affiliation{Instituto de F\'\i sica, Universidade de S\~ao Paulo\\
Caixa Postal 66318, 05315-970, S\~ao Paulo, SP, Brazil}%


\begin{abstract}

In this work, we develop a normal product algorithm suitable to the study of anisotropic field theories
in flat space, apply it to
construct the symmetries generators  and describe  how their possible anomalies may be found. In particular, we
discuss the dilatation anomaly in a scalar model with critical exponent $z=2$ in six spatial dimensions.

\end{abstract}
\maketitle


\section{Introduction}
Aiming the ultraviolet  improvement  of the perturbative series without the introduction of ghost degrees of freedom, field theories with higher spatial
derivatives have been proposed in the literature \cite{Horava1,Horava2,Anselmi1,Anselmi4,Anselmi2,Anselmi3,Visser}. The characteristic feature of these proposals is provided by the different scaling properties of  space and   time,
i.e., $x^i \rightarrow b x^i$ whereas $ t\rightarrow b^{z}t $. The smallest value of the critical exponent $z$ is one which corresponds to the
usual Lorentz symmetric field theories; higher values of $z$ furnishes models with better ultraviolet behavior at the expenses of breaking Lorentz invariance.
Many investigations  of theories with such anisotropy  have been reported, namely,  quantization  of gravitational models \cite{Henneaux,Restuccia,Blas,Orlando1,Pospelov,Bemfica}, applications to cosmology \cite{Kiritsis,Brandenberger,Calcagni,Saridakis,Mukohyama}, studies in Lorentz symmetry restoration and the renormalization group \cite{Iengo,Dhar,Dhar2,Iengo1,Gomes,Kikuchi} and other aspects of field theories \cite{Das,Orlando2,Alexandre1,Alexandre2,Chen,He,Nacir,Eune,Farakos,Petrov,Xue,Redigolo}.

 In a previous work \cite{Gomes} we studied  properties of the renormalization of these anisotropic field theories and analyzed the renormalization group flows of relevant parameters in various models with special emphasis on the possibility of Lorentz symmetry restoration at low energies. We found that, the restoration requires that the interactions be infrared stable and  some process of dimensional reduction must be furnished to treat the eventual
divergences appearing whenever the  higher derivative quadratic terms are eliminated.
 Besides the renormalization aspects,  it is natural  to
investigate the underlying symmetries and their possible anomalies. In the usual relativistic setting, special interest is devoted to the conformal and  chiral anomalies not only for their conceptual aspects but also for phenomenological applications (see\cite{Fujikawa} and references therein). For anisotropic space-time models these anomalies have been discussed in \cite{Adam}
for a scalar field coupled to a Ho\v rava-Lifshitz background and in \cite{Dhar,Bakas1,Bakas2} for the chiral anomaly of  Dirac fermions coupled to  gauge and   gravitational fields.
Pursuing these studies, in the  present work we generalize Zimmermann's normal product algorithm \cite{Zimmermann},  which for Lorentz symmetric theories has proved to be a very powerful tool \cite{Lowenstein}.
 We then  apply the formalism to the analysis
of the scale symmetry in a renormalizable $\varphi^4$ model with $z=2$ in six spatial dimensions where it is renormalizable.

Our work is organized as follows. In the section II we recall and extend the basics of the normal product algorithm which shall be used in the sequel. Thus, in section III we derive the conservation laws associated to space and time translations. Due to  the breaking
of Lorentz invariance inherent to theories with higher spatial derivatives, these two symmetries are considered separately. In section IV we examine the dilatation current and explicitly analyze the anomaly in its conservation law. A summary and additional remarks are presented in the Conclusions.
One appendix to study Noether's theorem appropriated to higher derivative theories is included.


\section{Normal Products in Anisotropic Field Theories}


Throughout this work we will consider the BPHZ renormalized anisotropic scalar field model, with $z=2$ in $d=6$ spatial dimensions, given by

\begin{equation}
\mathcal{L}=\frac12 (1+A)\partial_{0}\varphi \partial_{0}\varphi-\frac{(b^{2}+B)}{2} \partial_{i}\varphi \partial_{i}\varphi -
\frac{(a^{2}+C)}{2}\Delta\varphi \Delta\varphi-\frac{(m^2+D)}{2}\varphi^2-\frac{(\lambda+E)}{4!}\varphi^4,
\label{1.1}
\end{equation}
where $A$, $B$, $C$, $D$ and $E$ are finite counterterms fixed according to appropriated renormalization conditions,
$b^2$, $a^2$, $m^2$ and $\lambda$ are the original parameters in the Lagrangian and $\Delta$ denotes the spatial Laplacian
$\Delta\equiv \partial_i\partial_i$.

Before embarking into the analysis of the model (\ref{1.1}), it is useful to recall and generalize the BPHZ normal product
formalism \cite{Zimmermann,Lowenstein}. Thus, let $\mathcal{O}$ be some formal product of fields and their derivatives. Associated to the operator $\mathcal{O}$,
there exists an infinite set of normal products $N_{\delta}$, where $\delta$, the degree of the normal product,
must be an integer greater than or equal to the canonical dimension of the operator $\mathcal{O}$. More precisely,
\begin{equation}
\langle 0 |T N_{\delta}[ \mathcal{O}] X | 0\rangle\equiv \text{Finite Part of} \, \langle 0 |T  \mathcal{O} X | 0\rangle,
\label{1.2}
\end{equation}
where $X\equiv \prod_i \varphi_i(x_i)$ with $\varphi_{j}$ designating the field of type "$j$", either bosonic or fermionic.  The prescription to get the finite part can be described as follows.
We suppose that, without the normal product insertion, the Green's function are already renormalized, according to the BPHZ scheme. However,
in the presence of the formal product ${\cal O}$ there are additional divergences that need to be subtracted. Indeed, the superficial degree of divergence for a generic proper graph $G$, with $N_{B}$ and $N_{F}$ bosonic and fermionic external lines, is in the anisotropic case
\begin{equation}
d(G)=d+z -\mbox{Dim}[\varphi] N_B-\mbox{Dim}[\psi] N_F - \sum_{a}(d+z - \mbox{Dim}[V_a]),
\label{1.3}
\end{equation}
where the sum is over all vertices of $G$, the dimensions of the bosonic ($\varphi$) and fermionic ($\psi$) fields are
\begin{equation}
\mbox{Dim}[\varphi]= \frac{d-z}2,\qquad \qquad\mbox{Dim}[\psi]= \frac{d}2
\label{1.4}
\end{equation}
and Dim[$V_a $]= $D_{a}$+Dim[$\varphi$]$\nu^{B}_{a}$+Dim[$\psi$]$\nu^{F}_{a}$. $\nu^{B}_{a}$ and $\nu^{F}_{a}$ are the
number of bosonic and fermionic lines joining at the vertex $V_a$  and $D_{a}$ counts the number of momenta factors in the vertex, $z$ for a  time-like
and one for a space-like component.

 In a renormalizable model all vertices coming from the Lagrangian are assumed to have $\text{Dim}[V_a]=d+z$, whereas for the special vertex $V_{\cal O}$, associated to the operator ${\cal O}$, we make the replacement Dim[${\cal O}$] $\rightarrow \delta$. This designation is convenient although it implies that super-renormalizable vertices are going to be oversubtracted.  By taking into consideration these observations, the degree function  $\delta(\gamma)$ for  a generic proper graph $\gamma $ is defined to be
 \begin{equation}
\delta(\gamma)=\left\{
\begin{array}
[c]{cc}
\delta-\mbox{Dim}[\varphi] N_B-\mbox{Dim}[\psi] N_F,& \text{if}~~ V_{\mathcal{O}}~ \in~ \gamma\\
d+z-\mbox{Dim}[\varphi] N_B-\mbox{Dim}[\psi] N_F,& \text{if}~~ V_{\mathcal{O}}~ \in\!\!\!\!\!\slash~ \gamma.\\
\end{array}
\right.
\end{equation}
Finally,  the  finite part prescription consists in the application of the forest formula of the BPHZ scheme,
such that for a proper primitively divergent diagram $\gamma$ we must  apply the Taylor operator of degree $\delta(\gamma)$,
\begin{equation}
t^{\delta(\gamma)}I_{\gamma}\equiv\sum_{s=0}^{[\frac{\delta(\gamma)}{z}]}\frac{p_{0}^{s}}{s!}\frac{\partial^{s}{\phantom a}}{\partial p_{0}^{s}}\sum_{n=0}^{\delta(\gamma)-sz}\frac{{ p}_{i_{1}}\ldots{ p}_{i_{n}}}{n!}\frac{\partial\phantom a}{\partial{ p}_{i_{1}}}\ldots\frac{\partial\phantom a}{\partial{ p}_{i_{n}}}I_{\gamma},
\end{equation}
where $[x]$ is the greatest integer less than or equal to $ x $, $ p_{0}^{s} $ symbolically stands for the product of $ s $ time-like components of an independent  set of external momenta; ${ p}_{i}$ denotes the i-th space-like momentum (with the index of the component implicit) and all derivatives are computed at zero external momenta.

The normal products so defined  satisfy a set of rules that enable us to derive  Ward identities in a systematic way. To exemplify the general procedure let us consider the case of a partially conserved symmetry, i.e., a symmetry that formally holds when the breaking parameters vanish
(as the dilatation symmetry in (\ref{1.1})  when $b$ and $m$ go to zero).  In the normal product formalism, the conservation law associated with a  partially conserved continuous  symmetry  is obtained by taking as the generator of the transformations the  normal product  of the corresponding Noether's current.
One then studies the object
\begin{equation}
\partial_\mu \langle 0|TN[J^{\mu}](x) \prod \varphi(x_i)|0\rangle,
\end{equation}
where the degree of the normal product, for each component of the current $J^{\mu}$,  should be the  minimal  leading to a well defined expression.  As explained below, the derivative can be taken inside the normal product symbol by adequately increasing  the degree of the normal product. The equations of motion are afterwards applied producing  Dirac delta terms  characteristics of the symmetry  accompanied
by other terms which break the symmetry. Generally, the breaking terms are oversubtracted and to investigate their persistence
at the symmetry limit, Zimmermann identities, to be discussed shortly, are used to reduce the degrees of the normal products to the minimal ones what unveils  the anomalies.
To implement this algorithm the normal product formalism furnishes the following rules:

1. Differentiation rule. It can be   show that
\begin{equation}
\partial_0\langle 0 |T N_{\delta}[ \mathcal{O}] X | 0\rangle=\langle 0 |T N_{\delta+z}[\partial_0 \mathcal{O}] X | 0\rangle \qquad \mbox{and}\qquad
\partial_i\langle 0 |T N_{\delta}[ \mathcal{O}] X | 0\rangle=\langle 0 |T N_{\delta+1}[ \partial_i\mathcal{O}] X | 0\rangle.
\end{equation}
In fact, these results follow easily from the observation that the anisotropic Taylor operator of degree $\delta$ satisfy
$p^0 t_p^{\delta} f(p)= t_p^{\delta+z}(p^0 f(p))$ and $p^i t_p^{\delta} f(p)= t_p^{\delta+1}(p^i f(p))$.

2. Equation of Motion. It is the quantum version of the classical (Euler-Langrange) equation of motion  and  for the model (\ref{1.1}) it is given by
\begin{eqnarray}
&&\langle 0|{\rm T} N_{8}[\varphi(\partial_{0}^{2}-{b^{2}}\Delta+a^{2}\Delta^{2}+ m^{2})\varphi](x) X|0\rangle=
-A\langle 0|T N_8[\varphi \partial_0^2\varphi](x)X |0\rangle\nonumber\\ &+&B\langle 0|T N_8[\varphi \Delta\varphi](x)X |0\rangle
-C\langle 0|T N_8[\varphi \Delta^2\varphi](x)X |0\rangle-D\langle 0|T N_8[\varphi^2 ](x)X |0\rangle\nonumber\\
&-&\frac{(\lambda+E)}{3!}\langle 0|{\rm T}N_{8}[\varphi^{4}](x) X|0\rangle
-i\sum_{i=1}^{N}\delta(x-x_{i})\langle 0|{\rm T}X|0 \rangle.
\label{eom}
\end{eqnarray}
This expression may be derived by noting that in momentum space the operator applied on $\varphi$, in the left hand side of the above equation, is equal to $-i$ times the inverse of the free field propagator.


3. Zimmermann identities.
These are identities relating normal products of different degrees associated to the same formal product.
Using again the model (\ref{1.1}), let us consider some examples that will be used later on. First, to the operator $\varphi^2$, with dimension $\text{Dim}[\varphi^2]=4$,
we want to obtain the terms collectively denoted by ${\cal R}$ in the relation
\begin{equation}
\langle 0 |T N_8[\varphi^2]X| 0\rangle=\langle 0 |T N_4[\varphi^2]X| 0\rangle+{\cal R}.\label{ot}
\end{equation}
The degree functions for the $N_{4}$ and $N_{8}$ normal products  are respectively: $\delta_1(\gamma)=4-2N_{\gamma}$ and $\delta_2(\gamma)=8-2N_{\gamma}$. Thus:

a. For graphs with $N_{\gamma}=2$, $\delta_1(\gamma)=0$ and $\delta_2(\gamma)=4$. Notice first that the subtraction term without derivatives is present in both schemes so that it does not contribute to their difference. The subtractions with $\delta_2$ produce  the
following additional terms: $\langle 0 |T N_8[\varphi\Delta\varphi]X| 0\rangle$ and $\langle 0 |T N_8[\partial_i\varphi\partial_i\varphi]X| 0\rangle$
with second order derivatives, and $\langle 0 |T N_8[\varphi\partial_0^2\varphi]X| 0\rangle$, $\langle 0 |T N_8[\partial_0\varphi\partial_0\varphi]X| 0\rangle$,
$\langle 0 |T N_8[\Delta\varphi\Delta\varphi]X| 0\rangle$, $\langle 0 |T N_8[\varphi\Delta^2\varphi]X| 0\rangle$,
$\langle 0 |T N_8[\partial_i\partial_j\varphi\partial_i\partial_j\varphi]X| 0\rangle$ and
$\langle 0 |T N_8[\Delta\partial_i\varphi\partial_i\varphi]X| 0\rangle$ with
fourth order derivatives. Note that the terms of the second order derivatives are oversubtracted and can be still reduced, i.e., may be written in terms of minimally subtracted normal products.

b. For graphs with $N_{\gamma}=4$, $\delta_1(\gamma)=-4$ and $\delta_2(\gamma)=0$. The difference between the two normal products is just the subtraction term without derivatives associated to the scheme with $\delta_{2}$. This counterterm is therefore
 $\langle 0 |T N_8[\varphi^4]X| 0\rangle$.

 We could now collect all  the contributions described above to explicitly write  the terms denoted by ${\cal R}$ in (\ref{ot}) but,  instead, we will do that just for the integrated normal products, a procedure that avoids the dissemination of parameters and  that will be useful later on. In this situation, we obtain
 \begin{eqnarray}
\int \!dx^0 d^6x\langle 0 |T N_8[\varphi^2](x)X| 0\rangle&=&\int\! dx^0 d^6x\langle 0 |T \left \{N_4[\varphi^2](x)+
p N_8[\varphi\Delta\varphi](x)+sN_8[\partial_0\varphi\partial_0\varphi](x)\right. \nonumber\\&&
\left.+u N_8[\varphi\Delta^2\varphi](x)+v N_8[\varphi^4](x)\right \}X|0\rangle .
\label{zi1}
\end{eqnarray}
The coefficients $p,s,u,v$ are determined from normalization conditions, as we  will show shortly.
Incidentally, a simple dimensional analysis show us that: $\text{Dim}[p]=-2$ and
$\text{Dim}[s]=\text{Dim}[u]=\text{Dim}[v]=-4$. This detail  will be important in the study of the dilatation anomaly. After noticing that the term $\langle 0 |T N_8[\varphi\Delta\varphi]X| 0\rangle$
is oversubtracted   we can follow a similar reasoning as before to verify that
\begin{eqnarray}
\int \!dx^0 d^6x\langle 0 |T N_8[\varphi\Delta\varphi](x)X| 0\rangle&=&\int \!dx^0 d^6x\langle 0 |T\left \{ N_6[\varphi\Delta\varphi](x)+
g N_8[\partial_0\varphi\partial_0\varphi](x)\right.
\nonumber\\
&& + k N_8[\varphi\Delta^2\varphi](x) +\left. l ~ N_8[\varphi^4](x)\right \} X| 0\rangle.
\label{zi2}
\end{eqnarray}
From Eqs. (\ref{zi1}) and (\ref{zi2}) it follows now that
\begin{eqnarray}
\int \!dx^0 d^6x\langle 0 |T N_8[\varphi^2](x)X| 0\rangle&=&\int \!dx^0 d^6x\langle 0 |T\left  \{ N_4[\varphi^2](x)+
p  N_6[\varphi\Delta\varphi](x)+(s+pg)N_8[\partial_0\varphi\partial_0\varphi](x)\right. \nonumber\\&&+(u+pk) N_8[\varphi\Delta^2\varphi](x)+(v+p\,l)\left. N_8[\varphi^4](x)\right \}X| 0\rangle.
\label{zi4}
\end{eqnarray}

4. Normalization conditions. As the normal products are defined  by subtracting divergences through the application of Taylor operators, they satisfy simple normalization conditions as we will describe now.
Thus, if $D^M$ is a differential operator of degree $M\equiv z M_0+M_1$ in the momenta,
\begin{equation}
D^M=\frac{\partial^{M_0+M_1}}{\partial p_1^0\cdots\partial p_{M_0}^0 \partial p_{M_0+1}^{i_1}\cdots \partial p_{M_0+M_1}^{i_{M_1}}},
\end{equation}
then
\begin{equation}
D^M \Gamma^{(N)}_{\mathcal{O},\delta}(p,p_1,\cdots,p_N)\Big{|}_{p=p_1=\cdots=0}=\text{contribution of trivial diagram},\label{2}
\end{equation}
if $0\leq M\leq \delta-\mbox{Dim}[\varphi] N_B-\mbox{Dim}[\psi] N_F$, i.e., the order of differential operator must be lower than the degree of the Taylor operator.
$\Gamma^{(N)}_{\mathcal{O},\delta}(p,p_1,\cdots,p_N)$ denotes the (amputated) vertex function of $N=N_B+N_F$ points with the insertion of the normal product
$N_{\delta}[\mathcal{O}(x)]$:
\begin{eqnarray}
&&(2\pi)^{d+1}\delta(p+p_1+\cdots+p_N)\Gamma^{(N)}_{\mathcal{O}, \delta}(p,p_1,\cdots,p_N)\equiv\int dx^0 d^dx\prod_{i=1}^N dx^0_i d^dx_i
e^{i(px+\sum p_k x_k)}\nonumber\\
&\times&\langle 0|TN_{\delta}[\mathcal{O}(x)]\varphi_1(x_1)\cdots\varphi_N(x_N)|0\rangle^{prop}, \label{3}
\end{eqnarray}
where $prop$ means proper diagrams, i.e., 1PI diagrams.

We may use now the property (\ref{2}) to fix the parameters in the above  Zimmermann identities.
To determine $g$ in Eq. (\ref{zi2}), for example, we rewrite that equation for the vertex functions defined by (\ref{3}) with $N=2$, apply the operator of fourth order $D^4=\partial^2/\partial p_1^0\partial p_1^0$
and then put the  external momenta equal to zero. The only terms  that survive  are:  the term accompanying $g$, i.e., $\Gamma^{(2)}_{\partial_0\varphi\partial_0\varphi,8} $, which has only the trivial contribution; and the term involving the reduced normal product $\Gamma^{(2)}_{\varphi\Delta\varphi,6}$. This yields
\begin{equation}
g=\frac{1}{4}\frac{\partial^2}{\partial p_1^0\partial p_1^0}\int dx dx_1 dx_2 e^{ip_1 x_1+ip_2x_2}
\langle 0|T N_6[\varphi\Delta\varphi](x)\varphi(x_1)\varphi(x_2)|0\rangle^{prop}\Big{|}_{p_1=p_2=0},
\end{equation}
where we defined $dx\equiv dx_0 d^6x$. The others coefficients  can be determined similarly and we get

\begin{equation}
k=-\frac{1}{384}\frac{\partial^4}{\partial p_1^i\partial p_1^i\partial p_1^j\partial p_1^j}\int dx dx_1 dx_2 e^{ipx+ip_1 x_1+ip_2x_2}
\langle 0|T N_6[\varphi\Delta\varphi](x)\varphi(x_1)\varphi(x_2)|0\rangle^{prop}\Big{|}_{p=p_1=p_2=0}
\end{equation}
and
\begin{equation}
l=-\frac{1}{4!}\int dx \prod_{i=1}^4dx_i e^{ipx+i\sum_k p_k x_k}
\langle 0|T N_6[\varphi\Delta\varphi](x)\varphi(x_1)\varphi(x_2)\varphi(x_3)\varphi(x_4)|0\rangle^{prop}\Big{|}_{p=p_1=\cdots= p_4=0}.
\label{l}
\end{equation}
From these expressions we may verify that  all these  coefficients start contributing in second order in the coupling constant.
A similar analysis for Zimmermann identity (\ref{zi1}) show us that also $p,s,u,v$ all  start at second order
in the coupling constant. Thus, terms like $pg$ appearing in $s+pg$
are of higher order when  compared with $s$. This analysis of order, as well as the dimensions of parameters involved in
the Zimmermann identities, will be important in the determination of the dilatation anomaly.

Now, we are ready to discuss the conservation laws at quantum level and then investigate the existence of anomalies.
Let us start with space and time translations invariance.

\section{Energy-Momentum Tensor}

The classical energy-momentum tensor can be easily constructed from Noether's theorem for theories with higher derivatives (see appendix A)
applied to space-time translations $x^{\mu}\rightarrow x^{\mu}+\epsilon^{\mu}$.
Due to the asymmetric form of the Lagrangian, we need to distinguish  two cases:

a. Invariance under time translations. The components of the conserved current are
\begin{eqnarray}
&&\Theta_{00}=(1+A)\partial_0\varphi\partial_{0}\varphi-\mathcal{L},\label{2.1}\\ &&{\Theta}_{i0}=-(b^2+B)\partial_i\varphi\partial_{0}\varphi-(a^2+C)\Delta\varphi\partial_i\partial_{0}\varphi
+(a^2+C)(\partial_i\Delta\varphi) \partial_{0}\varphi,\nonumber
\end{eqnarray}
which satisfy $\partial_0\Theta_{00}+\partial_i{\Theta}_{i0}=0$ leading to the conservation of the energy $H=\int d^6x \,\Theta_{00}$.

b. Invariance under spatial translations. In this case the components of the conserved current are
\begin{eqnarray}
&& \Theta_{0j}=(1+A)\partial_0\varphi\partial_{j}\varphi,\label{2.2a}\\
&&{\Theta}_{ij}=-(b^2+B)\partial_i\varphi\partial_{j}\varphi-(a^2+C)\Delta\varphi\partial_i\partial_{j}\varphi
+(a^2+C)(\partial_i\Delta\varphi) \partial_{j}\varphi-\delta^{ij}\mathcal{L},
\end{eqnarray}
such that $\partial_0\Theta_{0j}+\partial_i{\Theta}_{ij}=0$ implies in the conservation of the $j$ component of the momentum: $\int d^6 x \, \Theta_{0j}$.

At the quantum level, to eliminate divergences, we should use normal products to define the energy and momentum operators.
Thus, notice that these components have different dimensions:
$\text{Dim}[\Theta_{00}]=8$, $\text{Dim}[\Theta_{0i}]=7$, $\text{Dim}[{\Theta}_{i0}]=9$ and
$\text{Dim}[{\Theta}_{ij}]=8$. So, we shall use the following normal
products $\langle 0| T N_8[\Theta_{00}]X |0\rangle$, $\langle 0| T N_7[\Theta_{0i}]X |0\rangle$,
$\langle 0| T N_9[{\Theta}_{i0}]X |0\rangle$ and $\langle 0| T N_8[{\Theta}_{ij}]X |0\rangle$.
According to the rules of normal products of the previous section, we get the analogous of the conservation equations
at the quantum level:
\begin{equation}
\partial_0\langle 0| T N_8[\Theta_{00}](x)X |0\rangle+\partial_i\langle 0| T N_9[{\Theta}_{i0}](x)X |0\rangle=
-i\sum_k\delta(x-x_k)\langle 0|T\partial_0\varphi(x)X_k |0\rangle
\label{emc1}
\end{equation}
and
\begin{equation}
\partial_0\langle 0| T N_7[\Theta_{0j}](x)X |0\rangle+\partial_i\langle 0| T N_8[{\Theta}_{ij}](x)X |0\rangle=
-i\sum_k\delta(x-x_k)\langle 0|T\partial_j\varphi(x)X_k |0\rangle,
\label{emc2}
\end{equation}
where $X_k$ is equal to the $X$ but with the field $\varphi(x_k)$ omitted.
By integrating over the space-time, the left hand sides of (\ref{emc1}) and (\ref{emc2}) vanish and we get
\begin{equation}
\sum_k\langle 0|T\partial_0\varphi(x_k)X_k |0\rangle=0~~~\text{and}~~~\sum_k\langle 0|T\partial_j\varphi(x_k)X_k |0\rangle=0.
\end{equation}
These Ward identities reflect the invariance of the Green functions under space and time translations.

\section{Dilatation Anomaly}

\subsection{Classical Analysis}

The crucial modification in anisotropic theories are just that of scale transformations, due to the anisotropic
scaling between space and time,

\begin{equation}
x_i\rightarrow e^\alpha x_i\,\,\,\,\text{and}\,\,\,\, x_0\rightarrow (e^\alpha)^z x_0.
\label{4.1}
\end{equation}
Under this scaling, the scalar field transform as $\varphi(x)\rightarrow \varphi^{\prime}(x^{\prime})=e^{-\alpha\text{Dim}[\varphi]}\varphi(x)$,
whose infinitesimal version is

\begin{equation}
\delta\varphi(x)\equiv\varphi^{\prime}(x)-\varphi(x)=\epsilon[\text{Dim}[\varphi]+z\,x^0\partial_0+x^i\partial_i]\varphi(x).
\label{4.2}
\end{equation}
We are particularly interested in the study of the consequences of this anisotropic scaling for the dilatation anomaly.
Let us consider the Lagrangian (\ref{1.1}) without  the  finite counterterms that for our (classical) purpose are not necessary,
\begin{equation}
\mathcal{L}=\frac12 \partial_{0}\varphi \partial_{0}\varphi-\frac{b^{2}}{2} \partial_{i}\varphi \partial_{i}\varphi -
\frac{a^{2}}{2}\Delta\varphi \Delta\varphi-\frac{m^2}{2}\varphi^2-\frac{\lambda}{4!}\varphi^4.
\end{equation}
The dimensional parameters $b^2$ and $m^2$  in this Lagrangian break the scale invariance as $\text{Dim}[m^2]=4$
and $\text{Dim}[b^2]=2$. On the other hand, when $b^2$ and $m^2\to 0$, we have a classical conserved current.
By splitting the Lagrangian in two parts, $\mathcal{L}=\mathcal{L}_{inv}+\mathcal{L}_{br}$, with
\begin{equation}
\mathcal{L}_{inv}\equiv\frac12 \partial_{0}\varphi \partial_{0}\varphi-
\frac{a^{2}}{2}\Delta\varphi \Delta\varphi-\frac{\lambda}{4!}\varphi^4
\end{equation}
and
\begin{equation}
\mathcal{L}_{br}\equiv-\frac{b^{2}}{2} \partial_{i}\varphi \partial_{i}\varphi -
\frac{m^2}{2}\varphi^2,
\end{equation}
we observe that, under the transformation (\ref{4.1}),
\begin{equation}
\delta\mathcal{L}_{inv}=\epsilon[2 \partial_0(x^0\mathcal{L}_{inv})+\partial_i(x^i\mathcal{L}_{inv})],
\end{equation}
is a total derivative. However, the same  does not happen with $\mathcal{L}_{br}$:
\begin{equation}
\delta\mathcal{L}_{br}=\epsilon[2 \partial_0(x^0\mathcal{L}_{br})+\partial_i(x^i\mathcal{L}_{br})]+\epsilon
(b^{2} \partial_i\varphi \partial_i\varphi +
2m^2\varphi^2).
\end{equation}
 From these results, the components of the Noether's current can be constructed,
\begin{equation}
J_0=2\varphi \partial_0\varphi+2 x^0\Theta_{00}+x^i\Theta_{0i}
\end{equation}
and
\begin{equation}
J^i=-3a^2\Delta\varphi\partial_i\varphi+2a^2(\partial_i\Delta\varphi)\varphi-2b^2\varphi\partial_i\varphi+
2x^0{\Theta}_{i0}+x^j{\Theta}_{ji},
\end{equation}
which satisfies

\begin{equation}
\partial_0 J^0+\partial_i J^i=b^2\partial_i\varphi\partial_i\varphi+2m^2\varphi^2,
\end{equation}
such that when $m^2$ and $b^2\rightarrow 0$, we have the conservation law.
On the other hand, in the quantum case we need to consider the subtraction scheme to provide
a precise meaning for the expressions involving operators at the same point,  giving rise to the anomaly.

\subsection{Quantum Analysis}

Now, let us apply the normal product method to the construction of the dilatation current. The dimension of components of the
current are: $\text{Dim}[J_0]=6$ and $\text{Dim}[J_i]=7$. So, we need to consider the normal products
$\langle 0|T N_6 [J_0](x)X |0\rangle$ and $\langle 0|T N_7 [J_i](x)X |0\rangle$. We are going to investigate the
conservation law
\begin{equation}
\partial_0 \langle 0|T N_6 [J_0]X |0\rangle+\partial_i\langle 0|T N_7 [J_i]X |0\rangle\,\, {\buildrel\,?\over=}
-i\sum_k\delta(x-x_k)\langle 0|T(\text{Dim}[\varphi]+2x^0\partial_0+x^i\partial_i)\varphi(x_k)X_k |0\rangle.
\label{q1}
\end{equation}
Let us rewrite the formal components of the current including the finite counterterms
\begin{equation}
J_0=d_a(1+A)\varphi \partial_0\varphi+2 x^0\Theta_{00}+x^i\Theta_{0i}
\end{equation}
and
\begin{equation}
J^i=-3(a^2+C)\Delta\varphi\partial_i\varphi+2(a^2+C)(\partial_i\Delta\varphi)\varphi-2(b^2+B)\varphi\partial_i\varphi+
2x^0{\Theta}_{i0}+x^j{\Theta}_{ji},
\end{equation}
where $d_a$ is the dimension of the scalar field. In lowest order of perturbation $d_a=2$, but we know
that in the quantum case it suffers radiative corrections, i.e., $d_a=2+\mathcal{O}(\lambda)$.
 In order to evaluate  (\ref{q1}) it is
convenient to use the identities
\begin{equation}
\langle 0|T N_8[x^i\partial_0\Theta_{0i}]X|0\rangle=x^i\partial_0\langle 0|T N_7[\Theta_{0i}]X|0\rangle
\end{equation}
and
\begin{equation}
\langle 0|T N_8[x^0\partial_i{\Theta}_{i0}]X|0\rangle=x^0\partial_i\langle 0|T N_9[{\Theta}_{i0}]X|0\rangle.
\end{equation}
Furthermore, employing the equations of the  energy and momentum conservation (\ref{emc1}) and (\ref{emc2}), and the equation of motion (\ref{eom}), we get
\begin{eqnarray}
&&\partial_0 \langle 0|T N_6 [J_0]X |0\rangle+\partial_i\langle 0|T N_7 [J_i]X |0\rangle=
(d_a-2)(1+A)\langle 0|TN_8[\partial_0\varphi\partial_0\varphi]X|0\rangle\nonumber\\&+&
(d_a-2)(b^2+B)\langle 0|TN_8[\varphi\Delta\varphi]X|0\rangle-(d_a-2)(a^2+C)\langle 0|TN_8[\varphi\Delta^2\varphi]X|0\rangle\nonumber\\
&-&(d_a-4)(m^2+D)\langle 0|TN_8[\varphi^2]X|0\rangle-(d_a-2)\frac{(\lambda+E)}{3!}\langle 0|TN_8[\varphi^4]X|0\rangle\nonumber\\&+&
(b^2+B)\langle 0|TN_8[\partial_i\varphi\partial_i\varphi]X|0\rangle
-i\sum_k\delta(x-x_k)\langle 0|T(d_a+2x^0\partial_0+x^i\partial_i)\varphi(x_k)X_k |0\rangle.
\end{eqnarray}
By integrating over the space and time, this expression reads
\begin{eqnarray}
&&i\sum_k\langle 0|T(d_a+2x^0\partial_0+x^i\partial_i)\varphi(x_k)X_k |0\rangle=\int dx^0d^6x\Big{\{}
(d_a-2)(1+A)\langle 0|TN_8[\partial_0\varphi\partial_0\varphi]X|0\rangle\nonumber\\&+&
(d_a-3)(b^2+B)\langle 0|TN_8[\varphi\Delta\varphi]X|0\rangle-(d_a-2)(a^2+C)\langle 0|TN_8[\varphi\Delta^2\varphi]X|0\rangle\nonumber\\
&-&(d_a-4)(m^2+D)\langle 0|TN_8[\varphi^2]X|0\rangle-(d_a-2)\frac{(\lambda+E)}{3!}\langle 0|TN_8[\varphi^4]X|0\rangle\Big{\}}.
\end{eqnarray}
Now, we may use Zimmermann identities (\ref{zi2}) and (\ref{zi4}) to reduce the normal products involving oversubtractions
$\langle 0|TN_8[\varphi\Delta\varphi]X|0\rangle$ and $\langle 0|TN_8[\varphi^2]X|0\rangle$. The final result is
\begin{eqnarray}
&&i\sum_k\langle 0|T(d_a+2x^0\partial_0+x^i\partial_i)\varphi(x_k)X_k |0\rangle=\int dx^0d^6x\Big{\{}
-(d_a-4)(m^2+D)\langle 0|TN_4[\varphi^2]X|0\rangle\nonumber\\&+&
[(d_a-3)(b^2+B)-(d_a-4)(m^2+D)p]\langle 0|TN_6[\varphi\Delta\varphi]X|0\rangle
\nonumber\\&+&[(d_a-2)(1+A)+(d_a-3)(b^2+B)g-(d_a-4)(m^2+D)\overline{s}]\langle 0|TN_8[\partial_0\varphi\partial_0\varphi]X|0\rangle\nonumber\\&+&
[-(d_a-2)(a^2+C)+(d_a-3)(b^2+B)k-(d_a-4)(m^2+D)\overline{u}]\langle 0|TN_8[\varphi\Delta^2\varphi]X|0\rangle\nonumber\\
&+&\Big{[}-(d_a-2)\frac{(\lambda+E)}{3!}+(d_a-3)(b^2+B)l-(d_a-4)(m^2+D)\overline{v}\Big{]}\langle 0|TN_8[\varphi^4]X|0\rangle\Big{\}},
\label{a}
\end{eqnarray}
where we have redefined: $\overline{s}\equiv s+pg$, $\overline{u}\equiv u+pk$ and $\overline{v}\equiv v+pl$.
To have  scale invariance symmetry of the Green functions, the coefficient of the three normal products $N_8$ must vanish independently,
which leads to the conditions
\begin{equation}
(d_a-2)(1+A)+(d_a-3)(b^2+B){g}-(d_a-4)(m^2+D)\overline{s}=0,
\label{cc1}
\end{equation}
\begin{equation}
-(d_a-2)(a^2+C)+(d_a-3)(b^2+B){k}-(d_a-4)(m^2+D)\overline{u}=0
\label{cc2}
\end{equation}
and
\begin{equation}
-(d_a-2)\frac{(\lambda+E)}{3!}+(d_a-3)(b^2+B)l-(d_a-4)(m^2+D)\overline{v}=0.
\label{cc3}
\end{equation}
Let us discuss the consistency of these equations. First of all, as noticed before, the terms $g$,
$\overline{s}$, $k$, $\overline{u}$, $l$ and $\overline{v}$ are of the
second order in the coupling constant.
So, by writing $d_a=2+d_a^{(1)}+d_a^{(2)}+\cdots$, where the superscript denotes the order
in the coupling constant, the equation (\ref{cc1}) show us that $d_a^{(1)}=0$. By writing
explicitly the order of each term, we obtain the
consistency condition at lowest non trivial order
\begin{equation}
d_a^{(2)}-b^2 {g}^{(2)}+2m^2\overline{s}^{(2)}=0,
\label{cc4}
\end{equation}
\begin{equation}
-a^2d_a^{(2)}-b^2k^{(2)}+2 m^2\overline{u}^{(2)}=0
\label{cc5}
\end{equation}
and
\begin{equation}
-b^2 l^{(2)}+2m^2\overline{v}^{(2)}=0.
\label{cc6}
\end{equation}
We will investigate this equations in the limit $b^2,~m^2\rightarrow 0$. Apparently, this equations are  solved by letting
$b^2,m^2=0$ and also  $d_a^{(2)}=0$. However, from the dimensional analysis of the parameters of the model, it follows that
  $g^{(2)}=\frac{\lambda^2}{b^2} \mathcal{F}(m^2/b^4,a^2)$, where
$\mathcal{F}$ is a dimensionless function of the parameters of the theory.
Thus, the product $b^2 g^{(2)} $ does not vanish when $b^2,~m^2\rightarrow 0$, with $m^2/b^4\rightarrow$ fixed.
The mass term behave analogously: $\overline{s}^{(2)}$ can be written as
$\overline{s}^{(2)}=\frac{\lambda^2}{m^2}\mathcal{G}(m^2/b^4,a^2)$ and
then the product $m^2\overline{s}^{(2)}$ does not vanish in the limit $b^2,~m^2\rightarrow 0$, with $m^2/b^4\rightarrow$ fixed.
The same  happens with the others terms in the equations (\ref{cc5}) and (\ref{cc6}): in the specified limit none of the terms containing $m^2$ or $b^{2}$ vanishes.
To see this explicitly, let us consider, for example, the condition (\ref{cc6}) in the limit $b^2,~m^2\rightarrow 0$, but $m^2/b^4\equiv c>0$.
The expression for $l^{(2)}$ is given by (\ref{l}), whereas $\overline{v}^{(2)}$ has the same form
that (\ref{l}), but with the replacing of the normal product $N_{6}[\varphi\Delta\varphi]$ by $N_4[\varphi^2]$. So,
 the left hand side of equation  (\ref{cc6}) is
\begin{equation}
\frac{6i^3\lambda^2}{4!}\int\frac{dk_0}{2\pi}\frac{d^6k}{(2\pi)^6}\frac{(b^2{\bf k}^2+2m^2)}{[k_0^2-b^2{\bf k}^2-a^2({\bf k}^2)^2-m^2+i\epsilon]^3}.
\end{equation}
By taking the limit above mentioned we get
\begin{equation}
\frac{6i^3\lambda^2}{4!}\int\frac{dk_0}{2\pi}\frac{d^6k}{(2\pi)^6}\frac{({\bf k}^2+2c)}{[k_0^2-{\bf k}^2-a^2({\bf k}^2)^2-c+i\epsilon]^3} =
\frac1{4!}\frac{3}{512\pi^3} \frac{ \lambda^2}{a^3},
\end{equation}
which is not zero for any finite $a$.
This shows that equation (\ref{cc6}) can not be satisfied,
and yields to the anomalous behavior of the quantum Ward identity.

\section{Conclusions}

In this work, we extended the method of the normal products to  anisotropic theories, what enables us to proceed
a systematic analysis of  Ward identities. In particular, we investigated the symmetry under space and
time translations in a $\varphi^4$ model and with $z=2$ in six spatial dimensions,
where the theory becomes renormalizable. Next, we studied the dilatation anomaly, caused by the
necessity of the introduction of a scale in the renormalization procedure. When compared with the
 relativistic case ($z=1,\,a=0,\, d=3$, with $b$  dimensionless),   in the anisotropic situation more  dimensional parameters are involved, associated with
the scale invariance breaking operators
$b^2\partial_i\varphi\partial_i\varphi$ and $m^2\varphi^2$ in the Lagrangian and hence in the divergence of the dilatation current.
In the context of the normal product formalism, these two operators appear oversubtracted, which means that they will produce
additional contributions to the anomaly. So, we may intuit that the anomaly is different from the relativistic case,
what was explicitly verified in (\ref{a}) and in the consistency conditions subsequently listed.

As expected, the divergence of the dilatation current is related to the renormalization group beta function. In fact,
a precise calculation show us that it is given essentially by  (\ref{cc6}),
 namely, $\beta(\lambda)=4!(-b^2 l^{(2)}+2m^2 \overline{v}^{(2)})=(3/512\pi^3 ) (\lambda^2/a^3)$.
This beta function was also obtained in \cite{Gomes}.
These aspects of the dilatation anomaly are in contrast
with the axial anomaly in gauge theories. As shown in
\cite{Dhar,Bakas1,Bakas2} by using the functional integral method, the form of the axial anomaly in the anisotropic situation is exactly the same as in its relativistic counterpart.
It has been argued that these differences are related to the ultraviolet character of the dilatation anomaly and
the infrared nature  of the axial anomaly \cite{Witten}.


\section{Acknowledgments}

This work was partially supported by  Conselho
Nacional de Desenvolvimento Cient\'{\i}fico e Tecnol\'ogico (CNPq) and
 Funda\c{c}\~ao de Amparo a Pesquisa do Estado de S\~ao Paulo (FAPESP).

\appendix

\section{ Noether's theorem for Lagrangians with third order derivatives}

For completeness, in this appendix we present a derivation of Noether's theorem appropriated to the cases studied in the text. Let us suppose that the basic field changes continuously from
$\varphi(x)$ to $\varphi(x)+\delta\varphi$ so that the Lagrangian density changes as
\begin{equation}
 \delta {\cal L} =\frac{\partial {\cal L}}{\partial \varphi}\delta \varphi+ \frac{\partial {\cal L}}{\partial\partial_{\mu} \varphi}\delta \partial_{\mu}\varphi+\frac{\partial {\cal L}}{\partial\partial_{\mu} \partial_{\nu}\varphi}\delta \partial_{\mu}\partial_{\nu}\varphi +\frac{\partial {\cal L}}{\partial\partial_{\mu} \partial_{\nu}\partial_{\rho}\varphi}\delta \partial_{\mu}\partial_{\nu}\partial_{\rho}\varphi.
 \end{equation}
Now, using the Euler-Lagrange equation of motion,
\begin{equation}
-\frac{\partial {\cal L}}{\partial\varphi}+ \partial_{\mu}\frac{\partial {\cal L}}{\partial\partial_{\mu}\varphi}-\partial_{\mu}\partial_{\nu}\frac{\partial {\cal L}}{\partial\partial_{\mu}\partial_{\nu}\varphi}+\partial_{\mu}\partial_{\nu}\partial_{\rho}\frac{\partial {\cal L}}{\partial\partial_{\mu}\partial_{\nu}\partial_{\rho}\varphi}=0,
\end{equation}
we get
\begin{equation}
\delta {\cal L} = \partial_{\mu}\left (\frac{\partial {\cal L}}{\partial\partial_{\mu}\varphi}\delta\varphi+\frac{\partial {\cal L}}{\partial\partial_{\mu}\partial_{\nu}\varphi}\der_{\nu}\delta \varphi\right ) +\left (\partial_{\mu}\partial_{\nu}\partial_{\rho}\frac{\partial {\cal L}}{\partial\partial_{\mu}\partial_{\nu}\partial_{\rho}\varphi} \right )\delta \varphi +\frac{\partial {\cal L}}{\partial\partial_{\mu} \partial_{\nu}\partial_{\rho}\varphi}\delta \partial_{\mu}\partial_{\nu}\partial_{\rho}\varphi,
\end{equation}
yielding
\begin{eqnarray}
\delta {\cal L} &=& \partial_{\mu}\left (\frac{\partial {\cal L}}{\partial\partial_{\mu}\varphi}\delta \varphi+\frac{\partial {\cal L}}{\partial\partial_{\mu}\partial_{\nu}\varphi}\der_{\nu}\delta \varphi\right ) +\partial_{\mu}\left (\partial_{\nu}\partial_{\rho}\frac{\partial {\cal L}}{\partial\partial_{\mu}\partial_{\nu}\partial_{\rho}\varphi} \delta \varphi \right )\nonumber\\
&-& \partial_{\nu}\left( \partial_{\rho}\frac{\partial {\cal L}}{\partial\partial_{\mu}\partial_{\nu}\partial_{\rho}\varphi}\partial_{\mu}\delta\varphi\right )+\partial_{\rho}\frac{\partial {\cal L}}{\partial\partial_{\mu}\partial_{\nu}\partial_{\rho}\varphi}\partial_{\nu}\partial_{\mu}\delta\varphi+
\frac{\partial {\cal L}}{\partial\partial_{\mu} \partial_{\nu}\partial_{\rho}\varphi}\delta \partial_{\mu}\partial_{\nu}\partial_{\rho}\varphi,
\end{eqnarray}
where $A\der_{\mu} B\equiv A\partial_{\mu}B-(\partial_{\mu}A)B$.
On the other hand, without using the equation of motion,
if we could show that the change in the Lagrangian is a total derivative $\delta\mathcal{L}=\partial_{\mu}S^{\mu}$,
we may therefore define a conserved current by
\begin{eqnarray}
J^{\mu}\equiv\frac{\partial {\cal L}}{\partial\partial_{\mu}\varphi}\delta \varphi+\left(\frac{\partial {\cal L}}{\partial\partial_{\mu}\partial_{\nu}\varphi}-\partial_{\rho}\frac{\partial {\cal L}}{\partial\partial_{\mu}\partial_{\nu}\partial_{\rho}\varphi}\right)\der_{\nu}\delta \varphi+\frac{\partial {\cal L}}{\partial\partial_{\mu}\partial_{\nu}\partial_{\rho}\varphi}\partial_{\nu}\partial_{\rho} \delta \varphi-S^{\mu}.
\end{eqnarray}
This result may be generalized. If the Lagrangian contains derivatives up to fourth order we arrive at
\begin{eqnarray}
J^{\mu}&=&\frac{\partial {\cal L}}{\partial\partial_{\mu}\varphi}\delta \varphi+\left(\frac{\partial {\cal L}}{\partial\partial_{\mu}\partial_{\nu}\varphi}-\partial_{\rho}\frac{\partial {\cal L}}{\partial\partial_{\mu}\partial_{\nu}\partial_{\rho}\varphi}+\partial_{\rho}\partial_{\sigma}\frac{\partial {\cal L}}{\partial\partial_{\mu}\partial_{\nu}\partial_{\rho}\partial_{\sigma}\varphi}\right)\der_{\nu}\delta \varphi+\frac{\partial {\cal L}}{\partial\partial_{\mu}\partial_{\nu}\partial_{\rho}\varphi}\partial_{\nu}\partial_{\rho} \delta \varphi\nonumber\\
&&+\frac{\partial {\cal L}}{\partial\partial_{\mu}\partial_{\nu}\partial_{\rho}\partial_{\sigma}\varphi} \der_{\nu}(\partial_{\rho}\partial_{\sigma}\delta \varphi)-S^{\mu}.
\end{eqnarray}



\begin{thebibliography}{55}
\bibitem{Horava1} P. Ho\v rava,{\it Quantum Gravity at a Lifshitz Point}, Phys. Rev. D79, 084008 (2009), arXiv:0901.3775. 	
\bibitem{Horava2} P. Ho\v rava, {\it Membranes at Quantum Criticality}, J. High Energy Phys. 0903, 020 (2009), arXiv:0812.4287.
\bibitem{Anselmi1} D. Anselmi and M. Halat, {\it  Renormalization of Lorentz violating theories}, Phys. Rev. D 76, 125011 (2007), arXiv:0707.2480.
\bibitem{Anselmi4} D. Anselmi, {\it Weighted scale invariant quantum field theories}, JHEP, 0802, 051 (2008), arXiv:0801.1216.
\bibitem{Anselmi2} D. Anselmi, {\it Weighted power counting and Lorentz violating gauge theories. I. General properties},
Ann. Phys. 324, 874 (2009), arXiv:0808.3470.

\bibitem{Anselmi3} D. Anselmi, {\it Weighted power counting and Lorentz violating gauge theories. II. Classification},
Ann. Phys. 324, 1058 (2009),  arXiv:0808.3474.
\bibitem{Visser} M. Visser, {\it  Lorentz symmetry breaking as a quantum field theory regulator}, Phys. Rev. D 80, 025011 (2009),
arXiv:0902.0590.
\bibitem{Henneaux} M. Henneaux, A. Kleinschmidt and G.  LucenaGomez,
{\it A dynamical inconsistency of Horava gravity}, Phys. Rev. D 81, 064002 (2010), arXiv:0912.0399.
\bibitem{Restuccia} J. Bellorin and A. Restuccia, {\it On the consistency of the Horava Theory}, arXiv:1004.0055.
\bibitem{Blas} D. Blas, O. Pujolas and S. Sibiryakov, {\it Consistent Extension of Horava Gravity},
Phys. Rev. Lett. 104 181302, (2010), arXiv:0909.3525.
\bibitem{Orlando1} D. Orlando and S. Reffert, {\it On the Renormalizability of Horava-Lifshitz-type Gravities},
Class. Quant. Grav. 26, 155021 (2009),  arXiv:0905.0301.	
\bibitem{Pospelov} M. Pospelov and Y. Shang, {\it On Lorentz violation in Horava-Lifshitz type theories}, arXiv:1010.5249.
\bibitem{Bemfica} F.S. Bemfica and M. Gomes, {\it  Fourth order spatial derivative gravity}, Phys. Rev. D 84, 084022 (2011), arXiv:1108.5979.	
\bibitem{Kiritsis} E. Kiritsis and G. Kofinas, {\it Horava-Lifshitz Cosmology}, Nucl. Phys. B 821:467-480, (2009), arXiv:0904.1334.
\bibitem{Brandenberger}R. Brandenberger, {\it Matter Bounce in Horava-Lifshitz Cosmology}, Phys. Rev. D 80, 043516 (2009), arXiv:0904.2835.
\bibitem{Calcagni} G. Calcagni, {\it Cosmology of the Lifshitz universe},  JHEP 0909, 112 (2009), arXiv:0904.0829.
\bibitem{Saridakis} E. N. Saridakis, {\it Horava-Lifshitz Dark Energy}, Eur. Phys. J. C 67 229, (2010), arXiv:0905.3532.
\bibitem{Mukohyama} S. Mukohyama, {\it Horava-Lifshitz Cosmology: A Review}, Class. Quant. Grav. 27:223101, (2010), arXiv:1007.5199.
\bibitem{Iengo}  R. Iengo, J. G. Russo and M. Serone, {\it Renormalization group in Lifshitz-type theories}, JHEP 0911, 020 (2009),
arXiv:0906.3477.
\bibitem{Dhar}A. Dhar, G. Mandal,  and S. R. Wadia, {\it Asymptotically free four-fermi theory in 4 dimensions at the z=3 Lifshitz-like fixed point}, Phys. Rev. D 80,  105018 (2009),	arXiv:0905.2928.
\bibitem{Dhar2} A. Dhar, G. Mandal and P. Nag, {\it Renormalization group flows in a Lifshitz-like four fermi model},  Phys. Rev. D 81, 085005, (2010),
arXiv:0911.5316.
\bibitem{Iengo1} R. Iengo and M. Serone, {\it A Simple UV-Completion of QED in 5D}, Phys. Rev. D 81,  125005 (2010), arXiv:1003.4430.
\bibitem{Gomes} P. R. S. Gomes and M. Gomes, {\it On Higher Spatial Derivative Field Theories}, arXiv:1107.6040.
\bibitem{Kikuchi} K. Kikuchi, {\it Restoration of Lorentz Symmetry for Lifshitz Type Scalar Theory}, arXiv:1111.6075.
\bibitem{Das} S. R. Das and G. Murthy, {\it $CP^{(N-1)}$ Models at a Lifshitz Point}, Phys. Rev. Lett. 104, 181601 (2010), arXiv:0906.3261.
\bibitem{Orlando2} D. Orlando and S. Reffert, {\it On the Perturbative Expansion around a Lifshitz Point}, Phys. Lett.  B 683, 62 (2010),
arXiv:0908.4429.
\bibitem{Alexandre1} J. Alexandre, {\it  Lifshitz-type Quantum Field Theories in Particle Physics}, Int. J. Mod. Phys. A 26:4523, (2011),
 arXiv:1109.5629.
\bibitem{Alexandre2} J. Alexandre, K. Farakos, P. Pasipoularides and A. Tsapalis, {\it Schwinger-Dyson approach for a Lifshitz-type Yukawa model},
Phys. Rev. D 81, 045002, (2010), arXiv:0909.3719.
\bibitem{Chen} B. Chen and Q. Huang, {\it  Field Theory at a Lifshitz Point}, Phys. Lett. B 683:108-113,2010, arXiv:0904.4565.
\bibitem{He} X. He, S. S. C. Law and R. R. Volkas, {\it Lifshitz theories with extra dimensions and 3+1-d Lorentz invariance}, Phys. Rev. D 84,
125017 (2011), arXiv:1107.3345.
\bibitem{Nacir} D. L. L. Nacir, F. D. Mazzitelli, L. G. Trombetta, {\it  Lifshitz scalar fields: one loop renormalization in curved backgrounds}, 
Phys. Rev. D 85, 024051 (2012), arXiv:1111.1662.
\bibitem{Eune} M. Eune, W. Kim and E. J. Son, {\it Effective potentials in the Lifshitz scalar field theory}, Phys. Lett. B 703:100-105, 2011,
arXiv:1105.5194.
\bibitem{Farakos} K. Farakos and D. Metaxas, {\it Symmetry breaking and restoration in Lifshitz type theories}, Phys. Lett. B 707, 562 (2012), arXiv:1109.0421.
\bibitem{Petrov} C.F. Farias, M. Gomes, J.R. Nascimento, A.Yu. Petrov and A.J. da Silva, {\it On the effective potential for Horava-Lifshitz-like theories}, arXiv:1112.2081.
\bibitem{Xue}  W. Xue, {\it Non-relativistic Supersymmetry}, arXiv:1008.5102.
\bibitem{Redigolo} D. Redigolo, {\it On Lorentz-Violating Supersymmetric Quantum Field Theories}, arXiv:1106.2035.
\bibitem{Fujikawa} K. Fujikawa and H. Suzuki, {\it Path Integrals and Quantum Anomalies}, Clarendon Press, Oxford 2004.

\bibitem{Adam} I. Adam, I. V. Melnikov and S. Theisen, {\it A Non-Relativistc Weyl Anomaly}, JHEP 0909:130, 2009, arXiv:0907.2156.

\bibitem{Bakas1} I. Bakas and D. Lust, {\it Axial anomalies of Lifshitz fermions}, Fortschr. Phys. 59 (2011) 937,
arXiv:1103.5693.
\bibitem{Bakas2}I. Bakas, {\it More on axial anomalies of Lifshitz fermions}, arXiv:1110.1332.
\bibitem{Zimmermann} W. Zimmermann,  in {\it Lectures on Elementary Particles and Quantum Field Theory}, edited by S. Deser,  
M. Grisaru, and H. Pendleton (MIT Press, Cambridge, 1970), 397.
\bibitem{Lowenstein} J. H. Lowenstein, {\it Normal product quantization of currents in Lagrangian field theory}, Phys. Rev. D4 2281, (1971).




\bibitem{Witten} L. Alvarez-Gaume and E. Witten, {\it Gravitational Anomalies}, Nucl. Phys. B 234:269, (1984).




\end{thebibliography}
\end{document}